\documentclass[%
 aip,
 amsmath,amssymb,
 reprint,%
]{revtex4-1}

\usepackage{graphicx}
\usepackage{dcolumn}
\usepackage{bm}

\usepackage[utf8]{inputenc}
\usepackage[T1]{fontenc}
\usepackage{mathptmx}

\begin{document}

\preprint{AIP/123-QED}

\title{Fabrication of Biaxial Hyperbolic Metamaterials with Oblique Angle Deposition}

\author{Changkee Hong}
 \altaffiliation{Department of Physics and Optical Engineering, Rose-Hulman Institute of Technology, 5500 Wabash Ave, Terre Haute, IN 47803, USA}

\author{Maarij Syed}
\altaffiliation{Department of Physics and Optical Engineering, Rose-Hulman Institute of Technology, 5500 Wabash Ave, Terre Haute, IN 47803, USA}

\author{Azad Siahmakoun}
\altaffiliation{Department of Physics and Optical Engineering, Rose-Hulman Institute of Technology, 5500 Wabash Ave, Terre Haute, IN 47803, USA}

\author{Hossein Alisafaee}
\altaffiliation{Department of Physics and Optical Engineering, Rose-Hulman Institute of Technology, 5500 Wabash Ave, Terre Haute, IN 47803, USA}

 \email{alisafae@rose-hulman.edu}





\begin{abstract}
We have demonstrated a versatile method for fabrication of biaxial hyperbolic metamaterials (HMMs) using layered structures consisting of titanium dioxide (TiO\textsubscript{2}) and copper (Cu).
In order to enable the biaxial property, an oblique angle deposition technique is applied to deposit the dielectric layer in the form of columnar nanostructures.
We have experimentally characterized the biaxial dispersion using variable angle spectroscopic ellipsometry measurements in the wavelength range 400nm to 900nm.
A noticeable difference between the in-plane permittivity components is observed which has proved to be immune from fabrication errors.
A biaxial HMM fabricated according to our method provides dual epsilon-near-zero properties at visible wavelengths which are separated over a bandwidth of 7nm.
The experimental characterization results of the fabricated biaxial HMM have been in good agreement with the predictions of effective medium approximation.
Our proposed method may be beneficial to accelerate the realization of new devices in nanophotonics applications.

\end{abstract}



\maketitle


Hyperbolic metamaterials (HMMs) are subwavelength engineered structures which exhibit a hyperbolic dispersion that provides metallic or dielectric characteristics in orthogonal directions~\cite{ferrari2015hyperbolic}.
The HMMs have an unbounded hyperboloid isofrequency surface which defines the anisotropic refractive index of the extraordinary waves, and supports arbitrarily large wavevectors to retain a propagating nature~\cite{ballantine2014conical,novotny2012principles}.
Conventional HMMs are made of either a subwavelength stack of alternating metallic and dielectric layers or a lattice of metallic nanowires embedded in a dielectric host.
Most recently, HMMs have been attracting significant attentions due to their unique role in nanophotonics applications.
Nanowire-based broadband HMMs have been proposed for nanophotonic chips and metamaterial-based flat lenses~\cite{Zang:19}.
Control of the spontaneous emission rate of dipole emitters has been investigated in HMMs~\cite{WangOL:19,lu2014enhancing} with applications in quantum optics, and on-chip quantum computing.
Photonic spin Hall effect has been experimentally demonstrated in a multilayer HMM at visible wavelengths~\cite{Takayama:18}.
Semiconductor HMMs~\cite{Sohr:19}, based on Si:InAs/AlSb, have been proposed for thermophotovoltaics and thermal emission management. 
HMM waveguides have also been designed for on-chip manipulation of light polarization~\cite{Makarova:17}.
The extent of applications exceeds beyond nanophotonics realm, covering topics such as Unruh effect~\cite{Smolyaninov:19}, magneto-optical effects~\cite{Kolmychek:18}, generation and focusing of terahertz photons~\cite{Popov:17,Kannegulla:16}, hyperlens~\cite{jacob2006optical}, and nanolithography~\cite{casse2010super}.
To date, most of the experimental investigations have been focused on uniaxial HMMs due to simplicity of fabrication.

\begin{figure}[t]
    \centering
    \includegraphics[width=1.0\columnwidth]{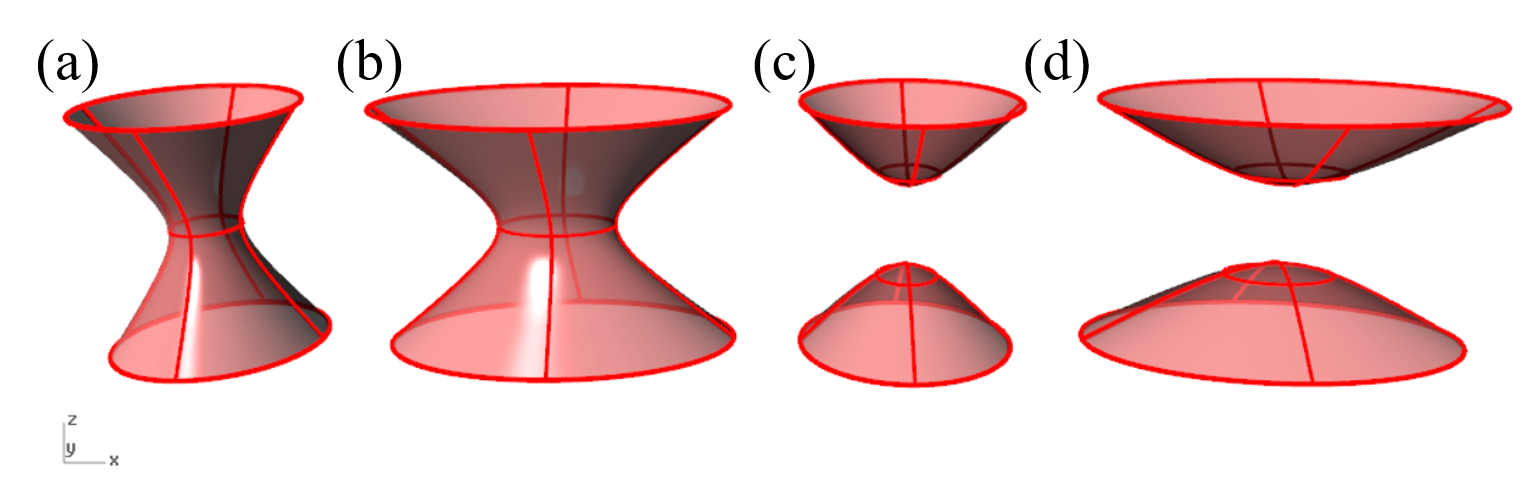}
    \caption{Representations of isofrequency surfaces for uniaxial (a,c) and biaxial (b,d) hyperbolic metamaterials.}
    \label{fig:isosurfs}
\end{figure}

Biaxial HMMs (BHMMs) exhibit an additional feature which is the asymmetry in the hyperbolic dispersion.
The normalized relative permittivity tensor of materials is described as:
\begin{equation}
\varepsilon =
  \left[ {\begin{array}{ccc}
   \varepsilon_{xx} & 0 & 0 \\
   0 & \varepsilon_{yy} & 0 \\
   0 & 0 & \varepsilon_{zz} \\
  \end{array} } \right].
\end{equation}
For the case of uniaxial HMMs in a layered configuration ~\cite{song2018biaxial}:
\begin{equation}
  \left\{
    \begin{array}{l}
      ~~~~~\varepsilon_{xx}=\varepsilon_{yy}\neq\varepsilon_{zz},\\
      \varepsilon_{xx}<0,\; \varepsilon_{yy}<0,\; \varepsilon_{zz}>0,
    \end{array}
  \right.
\end{equation}
which results in the isofrequency surface shown in Fig.~\ref{fig:isosurfs}a. 
Reversing the sign of permittivity components, another isofrequency surface can be obtained as in Fig.~\ref{fig:isosurfs}c. 
In the case of BHMMs, following:  
\begin{equation}
  \left\{
    \begin{array}{l}
      ~~~~~\varepsilon_{xx}\neq\varepsilon_{yy}\neq\varepsilon_{zz},\\
      \varepsilon_{xx}<0,\; \varepsilon_{yy}<0,\; \varepsilon_{zz}>0,
    \end{array}
  \right.
  \label{eq:bhmm}
\end{equation}
the existence of an additional birefringency results in an asymmetry in the isofrequency surfaces as shown in Figs.~\ref{fig:isosurfs}b and ~\ref{fig:isosurfs}d.
Asymmetric hyperbolic dispersion is attractive for applications in twisting light by conversion of Hermite-Gaussian beams into vortex beams carrying an orbital angular momentum~\cite{sun2013twisting}. 
Promising  applications  in  quantum  nanophotonics have been also enabled by the discovery of a new kind of surface wave on biaxial hyperbolic metamaterial, with elliptically polarized state that can be utilized for the spin-controllable excitation of surface waves, opening a gateway towards integrated photonic circuits with reconfigurable functionalities~\cite{gao2015chiral}.
However, most of the literature to date has only been devoted to theoretical predictions or full-field finite element simulations for such interesting applications, which indicates that the major challenge remains in the fabrication of BHMMs.  

\begin{figure}[t]
    \centering
    \includegraphics[width=.6\columnwidth]{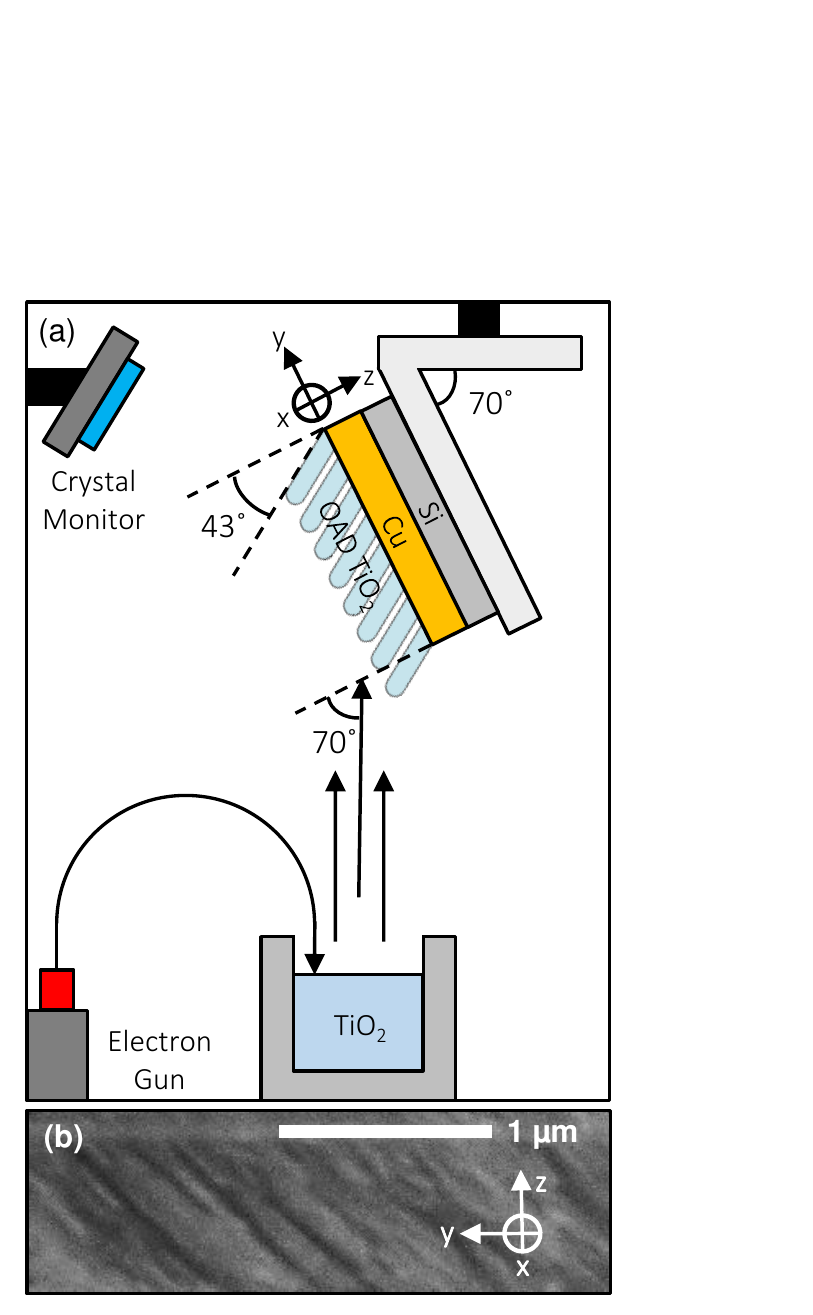}
    \caption{(a) Schematic of oblique angle deposition of TiO$_2$ in a physical vapor deposition system with substrate placed at an angle of $70^\circ$ resulting in growth of columnar structures at $43^\circ$ with respect to the substrate's normal. (b) Scanning Electron Microscope image of columnar nanostructures in OAD TiO$_2$.}
    \label{fig:pvd}
\end{figure}

Biaxial property in HMMs can be incorporated with addition of a birefringent element as the building block of the structure.
Black phosphor has been proposed theoretically as a candidate material for this type of BHMMs~\cite{song2018biaxial}, however the extent of the proposed study has been confined only to theoretical predictions.
In order to address the aforementioned challenge, we provide the details of our recent method for fabrication of a BHMM based on regularly accessible dielectric materials and metals.
We have exploited an approach that has demonstrated success in fabrication of birefringent photonic crystals with polarization-degenerate photonic band gaps~\cite{ordouie2018ultracompact}.
The technique is based on a simple but fundamental alternation in the method of deposition in fabrication of dielectric materials, which results in a birefringent thin film.
The method involves oblique angle deposition (OAD) of the dielectric material using physical vapor deposition (Fig.~\ref{fig:pvd}).
As it will be demonstrated, the OAD technique substantially eases the achievement of fabricating practical biaxial HMMs.

Oblique angle deposition (OAD) method allows fabrication of nanostructured columnar architectures.
In the OAD process, the substrate is tilted with respect to normal deposition direction~\cite{hawkeye2014glancing}.
With the start of the deposition, the atomic vapor is condensed on the substrate and forms microscopic nuclei.
As deposition continues, the vapor cannot be condensed immediately behind the nuclei, hence a small shadow region forms behind the nuclei, which eventually causes growth of nanocolumns with voids among them.
Such columnar structures demonstrate the birefringence in the plane of the layer.

The optical response of the discontinuous media composed of metal and dielectric could be estimated through effective medium approximation (EMA) theory~\cite{ferrari2015hyperbolic}.
We selected regularly accessible materials for the fabrication of BHHMs. 
Titanium dioxide (TiO\textsubscript{2}) was utilized as the dielectric to achieve the birefringent dielectric layer, and copper (Cu) was utilized at subwavelength scale in order to add metallic response.
The resulting composition presumably provides a biaxial hyperbolic dispersion according to theoretical predictions of EMA as described in the following.
For uniaxial HMMs, the permittivity tensor of dielectric is $\varepsilon_{D}$ and the permittivity of metal layer is $\varepsilon_{M}$.
Then, the components of the effective permittivity tensor in x, y, z directions can be expressed as~\cite{ferrari2015hyperbolic}:
\begin{equation}
\varepsilon_{xx} = \varepsilon_{yy} = \frac{\varepsilon_{M}\cdot p+\varepsilon_{D}}{p+1},
\label{eq:uni_exx}
\end{equation}
\begin{equation}
\varepsilon_{zz} = \frac{p+1}{\frac{p}{\varepsilon_{M}}+\frac{1}{\varepsilon_{D}}},
\end{equation}
where $p$ is the thickness ratio (fill factor) between metal and dielectric ($p = {d_{M}}/{d_{D}}$).
Accordingly, for the biaxial HMM, the components of the permittivity tensor of OAD dielectric in x, y, z directions are $\varepsilon_{OADx}$, $\varepsilon_{OADy}$, $\varepsilon_{OADz}$, and the thickness is $d_{OAD}$~\cite{song2018biaxial}. The permittivity of metal layer is $\varepsilon_{Metal}$, and the thickness is $d_{Metal}$.
The components of the effective permittivity tensor in x, y, z directions can be expressed as ~\cite{poddubny2013hyperbolic,wood2006directed,song2018biaxial}:
\begin{equation}
\varepsilon_{xx} =  \frac{\varepsilon_{Metal}\cdot p+\varepsilon_{OADx}}{p+1},
\end{equation}
\begin{equation}
\varepsilon_{yy} =  \frac{\varepsilon_{Metal}\cdot p+\varepsilon_{OADy}}{p+1},
\end{equation}
\begin{equation}
\varepsilon_{zz} =  \frac{p+1}{\frac{p}{\varepsilon_{Metal}}+\frac{1}{\varepsilon_{OADz}}}.
\label{eq:bi_ezz}
\end{equation}

\begin{figure}[t]
    \centering
    \includegraphics[width=1.0\columnwidth]{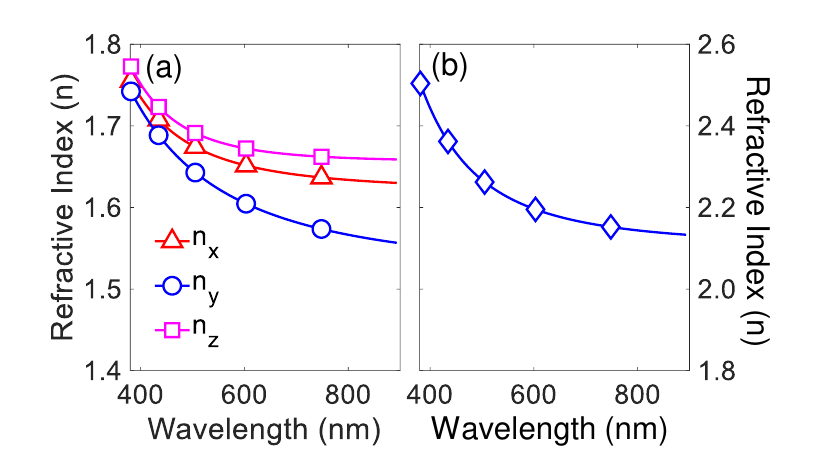}
        \caption{Measured index of refraction for (a) obliquely deposited TiO$_2$, (b) normally deposited TiO$_2$. Measurements are performed using variable angle spectroscopic ellipsometry at 65$^\circ$, 70$^\circ$, and 75$^\circ$ in the reflection mode.}
        \label{fig:oadindex}
\end{figure}

The EMA approach enables calculation and comparison of the optical responses between the uniaxial and biaxial HMMs.
For this purpose, the index of refraction for the dielectric deposited with OAD technique is required.
Therefore, a single layer of TiO\textsubscript{2} was deposited on a 4-inch silicon wafer (p-type Si:B [100] University Wafers) using electron beam PVD system (PVD 75, Kurt J. Lesker).
The wafer was placed at an angle of $70^\circ$ (see Fig.~\ref{fig:pvd}).
It is empirically known that the column angle of OAD TiO\textsubscript{2} is 43$^{\circ}$ when the deposition angle is 70$^{\circ}$~\cite{hodgkinson1998empirical}.
The single layer was then analyzed via an $\alpha$-SE J. A. Woollam spectroscopic ellipsometry system. 
The thickness of the OAD TiO\textsubscript{2} layer was determined in ellipsometry using Cauchy model for x, y, z directions, as 51.7nm $\pm$ 0.16nm with MSE of 3.14.
Figure~\ref{fig:oadindex}a shows the measured index of refraction for OAD deposited TiO$_2$.
As it was predictable, the index of this material in all directions is smaller than the normally deposited material which is shown in Fig.~\ref{fig:oadindex}b.

\begin{figure}[b]
    \centering
    \includegraphics[width=1.0\columnwidth]{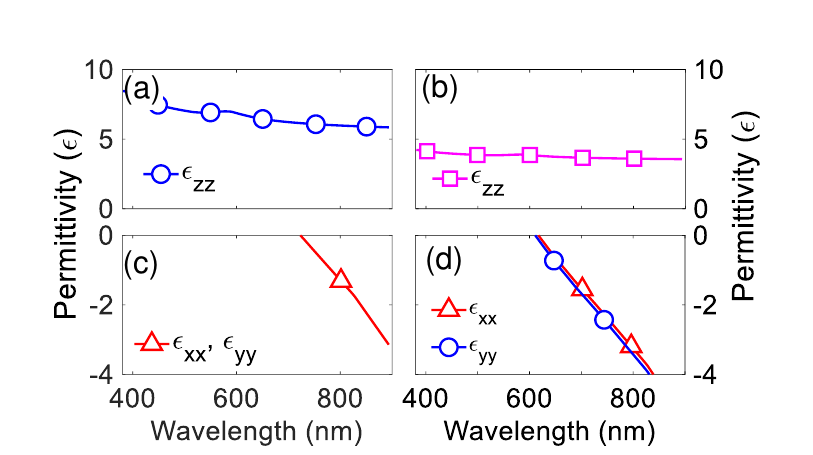}
        \caption{Real part of the components of permittivity tensor for (a,c) uniaxial HMM, and (b,d) biaxial HMM. Calculations performed using effective medium approximation with a fill factor of 0.30 for the ratio of metal (Cu) to dielectric (TiO$_2$). }
        \label{fig:hmmvsbhmm_theory}
\end{figure}

\begin{figure}[t]
    \centering
    \includegraphics[width=1.0\columnwidth]{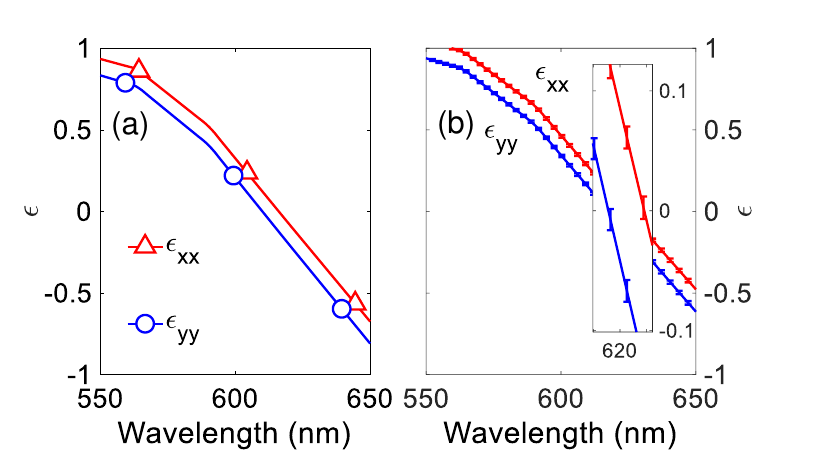}
        \caption{Real part of permittivity tensor for the in-plane components. (a) Theoretical calculations using effective medium approximation, (b) experimental measurements using spectroscopic ellipsometry. The inset shows a zoom into epsilon-near-zero region.}
        \label{fig:bhmm_mesur}
\end{figure}

Having extracted the indices of the materials, the EMA calculations, Eqs.~\ref{eq:uni_exx}-\ref{eq:bi_ezz}, result in the theoretical prediction of both uniaxial and biaxial HMMs made out of TiO$_2$ and Cu. 
A fill factor of $p=0.30$ was chosen since the portion of OAD TiO\textsubscript{2} must be large enough within one spatial period to obtain a discernible biaxial property of HMMs. 
The results are shown in Fig.~\ref{fig:hmmvsbhmm_theory}.
As expected, the z component of permittivity is positive in both the uniaxial (Fig.~\ref{fig:hmmvsbhmm_theory}a) and biaxial (Fig.~\ref{fig:hmmvsbhmm_theory}b) HMMs with the latter having a smaller response comparatively caused by reduced index of refraction in OAD TiO$_2$.
The in-plane permittivities ($\varepsilon_{xx}$ and $\varepsilon_{yy})$ are identical for the uniaxial HMM (Fig.~\ref{fig:hmmvsbhmm_theory}c) with a cross-over wavelength of 723nm at which the permittivity is zero and then it becomes negative at all the longer wavelengths.
This band is where the uniaxial HMM property is manifested (Fig.~\ref{fig:isosurfs}a).
On the other hand, the BHMM provides a blue-shift in the cross-over wavelength for $\varepsilon_{xx}$ and $\varepsilon_{yy}$ (Fig.~\ref{fig:hmmvsbhmm_theory}d), which respectively occur at 616nm and 609nm (also see Fig.~\ref{fig:bhmm_mesur}a).
Therefore, it can be deduced that at all the wavelengths longer than 616nm a biaxial behaviour is expected from the HMM, satisfying Eq.~\ref{eq:bhmm} (Fig.~\ref{fig:isosurfs}c). 
The 7-nm band between 609nm and 616nm is an  interesting region as it exhibits the properties of biaxial materials but with two responses corresponding a hyperbolic and an ellipsoid isofrequency surface in y and x directions, respectively.


Fabrication of the BHMM described above has been achieved using OAD technique.
For this purpose, first 15nm of Cu was deposited normally on a 4-inch silicon wafer using electron beam PVD system (PVD75, Kurt J. Lesker) at a rate of 0.5 \AA/s at $5\times10^{-5}$ Torr. 
With known optical  constants~\cite{palik1998handbook}, ellipsometric measurements have shown that the Cu layer was fabricated to a thickness of 15.5 $\pm$ 0.04nm (MSE = 17.82).
In the next step, the substrate was tilted to $70^\circ$, and then a very similar OAD layer of TiO$_2$ (as described earlier) was deposited on top of the Cu layer as a rate of 0.1\AA/s (Fig.~\ref{fig:pvd}).

In order to experimentally verify the biaxial hyperbolic response, variable angle spectroscopic ellipsometry (VASE) was used to extract the permittivity tensor over the entire visible spectrum.
After deposition of Cu and OAD TiO\textsubscript{2}, the total thickness of BHMM was measured with a profilometer (D-500, KLA Tencor), resulting in an average of 76.9 $\pm$ 5.51nm. 
Then, the total thickness of the fabricated BHMM was measured in VASE.
To extract all the components of the permittivity tensor, we employed a low-MSE extraction procedure~\cite{dilts2019low} with the following modifications:
A biaxial model with Bruggeman effective medium approximation in x, y, z directions was applied, where the previously obtained OAD TiO$\textsubscript{2}$ optical parameters (the data from Fig.~\ref{fig:oadindex}) as well as the optical constants of Cu with an EMA percentage of 21.1\% was employed.
Depolarization factors were set to 0 in x and y directions, and to 1 in z direction based on geometrical orientations of the layers. 
The fitting model resulted in a BHMM thickness of 72.9 $\pm$ 0.28nm with MSE of 18.3, leading to an effective thickness of 57.4nm for OAD TiO$_2$ ($p = 0.27$).
Therefore, the fabricated BHMM is remarkably similar to the designed BHMM.
These measurements were repeated and reproduced eight times in order to establish a valid statistical population and provide an estimate of the error.
In case the error of the measurements were larger than the difference in permittivity components, the fabricated BHMM would have been deemed futile. 
Thankfully, the measurements of the BHMM sample proved to have significant repeatability and reproducibility, with an error less than the difference between the permittivity components, clearly shown in Fig.~\ref{fig:bhmm_mesur}b.
Therefore, the fabrication of a practical BHMM has been successfully demonstrated.
It is also observed that the experimental results obtained for the fabricated biaxial HMM is in good agreement with the predictions of the effective medium approximation (Fig.~\ref{fig:bhmm_mesur}a). 
In the fabricated BHMM, a 7-nm band was also observed between the cross-over wavelengths for x and y components of permittivity corresponding to 617nm and 624nm, respectively.
The observed band shift is most probably due to factors such as a small deviation in the fill factor, a slight gradient in the thickness of the fabricated OAD TiO$_2$, and minor misalignment during the ellipsometry measurement trials.

In conclusion, we have successfully fabricated a biaxial hyperbolic metamaterial using oblique deposition of a dielectric material on a metallic layer at subwavelength scale.
To the best of our knowledge, this is the first time a biaxial hyperbolic metamaterial has been fabricated with this approach.
The fabricated device was characterized using variable angle spectroscopic ellipsometry for extraction of the permittivity tensor.
The results of experimental characterization are significantly close to the predictions by effective medium approximation.
Our approach in achieving biaxial HMMs provides an advantageous path for rapid fabrication of such devices and implementation of their exotic properties in a variety of applications in nanophotonics.

\begin{acknowledgments}
This work was performed in part at the MiNDS facility at Rose-Hulman Institute of Technology with the assistance of Brian Fair and Dr. Scott Kirkpatrick.
\end{acknowledgments}






\bibliography{OLBHMM1122}


\end{document}